\documentclass[Showpacs,preprintnumbers,amsmath,twocolumn]{revtex4}
\usepackage{graphicx}
\usepackage{dcolumn}
\usepackage{bm}
\usepackage{amssymb}
\usepackage{amsmath}
\usepackage{epsf}
\usepackage{subfigure}
\usepackage{epstopdf}%
\usepackage{amsfonts}

\def\tp{{t^\prime}}

\def\ip{{i^\prime}}
\def\jp{{j^\prime}}

\def\sp{{s^\prime}}

\begin{document}

\title{Coherent Stimulated X-ray Raman Spectroscopy; Attosecond Extension of RIXS}
\author{Upendra Harbola$^1$ and Shaul Mukamel$^2$}
\affiliation{$^1$Department of Chemistry and Biochemistry, University of California, San Diego,
California 92093-0340, United States\\
$^2$Department of Chemistry, University of California,
Irvine, California 92697-2025, United States}
\date{\today}

\begin{abstract}
Spontaneous and stimulated resonant inelastic X-ray Raman scattering signals are calculated using the Keldysh-Schwinger
closed-time path loop and expressed as overlaps of Doorway and Window electron-hole wavepackets. These are recast in terms of the one-particle Green's functions and expansion coefficients of configuration interaction singles for valence excitations, which can be obtained from standard electronic structure codes. Calculation for many-body states of ground and core-excited system is avoided.

\end{abstract}

\maketitle
Resonant nonlinear spectroscopy in the X-ray regime may become
possible by new bright ultra-fast sources
\cite{markusScience2001,serviceScience2002,science07,23,dorchies}. The theoretical
formulation of nonlinear spectroscopy with attosecond X-ray pulses
is of considerable interest
\cite{shaul-PRB,felicissimoJCP2005,IgorPRL,tanakaPRL2002}. The
picosecond optical pump/x-ray probe technique has been used to study
photophysical and photochemical molecular processes \cite{24,25,28,santra}.
An all-X-ray pump-probe experiment with attosecond
X-ray pulses has been proposed in Ref. \cite{IgorPRL}. The pump
pulse interacts with the system to create a valence excited state
wave-packet which evolves for a controlled delay time $\tau$ when a second
probe pulse interacts with it. $\tau$ is not limited by the core-hole
life-time. The dependence of this coherent stimulated X-ray Raman signal (CXRS) on the delay
time $\tau$ carries information about valence
excited state dynamics. The pump-probe signal may be recast in the
Doorway/Window representation of optical nonlinear spectroscopy
\cite{shaul-book}. In Ref. \cite{IgorPRL} these were computed within
the equivalent core single Slater determinant approximation. Here we express them
using the single-body Green's functions. Thus avoiding the explicit
computation of the many electron core excitations.

Green's functions have been
extensively used to study X-ray absorbtion fine structure (XAFS)
\cite{ashleyPRB1975,rehrCCR2005,soininenPRB2005,ankudinovPRB1998,rehr,feng08}.
The formalism is well developed and incorporates intrinsic
and extrinsic losses \cite{rehr} coming from the many-body
interactions (electron-phonon, electron-hole pairs etc.). These
provide a high level, yet practical, approach that goes beyond the
density-functional theory considered in Ref. \cite{IgorPRL}. By invoking
the "sudden approximation" whereby a core-hole is created
and destroyed instantaneously we express the signal
in term of one-particle Green's functions which depend on the
core-hole parametrically\cite{rehr}.

We compare CXRS with resonance inelastic X-ray Raman scattering (RIXS) which is a common
frequency-domain technique used in the study of core excited states
in solids and molecules\cite{agrenPhysRep1999,kotani,26,hamalainen}.
The RIXS signal can be described by the Kramers-Heisenberg expression
\cite{kotani,hikosakaPRL2008} as is done for valence excitations in the optical regime
\cite{shaul-book}.
\begin{eqnarray}
\label{1}
S_{RIXS}(\omega_1,\omega_2) = \sum_{ac}|A_{ca}(\omega_1)|^2 \delta(\omega_1-\omega_2-\omega_{ca})
\end{eqnarray}
with the transition amplitude
\begin{eqnarray}
\label{2}
A_{ca}(\omega_1) = \sum_{e} \frac{B_{ce}B_{ea}}{\omega_1-\omega_{ea}+ i\Gamma_{ea}}
\end{eqnarray}
here $\omega_1$ and $\omega_2$ are the incoming and scattered
photon frequencies, $a$ and $c$ denote the valence $N$-electron
ground state and singly excited states and $e$ is the excited state with one core-hole and $N+1$
valence electrons. $B_{ea}$ and $B_{ce}$ are matrix
elements of dipole operator.

Displaying this signal vs. $\omega_1-\omega_2$ reveals the valence
transitions ($\omega_{ca}$). The core-hole resonances $\omega_{ea}$
in the transition amplitude $A_{ca}$ are typically much broader and
less resolved due to the large core-hole lifetime contribution to $\Gamma_{ea}$
is $\sim 0.05$ eV \cite{hikosakaPRL2008}. CXRS is a closely related technique, performed with a pair of
attosecond pulses.  We show how that how the signal can be recast in a form
resembling Eq.(\ref{1}), making it a natural time-domain extension
of RIXS. Varying the envelopes of both pulses and their delay
offer a much higher degree of control of the signal. Both signals
can be described as a valence electron-hole wavepacket (a doorway state)
prepared by the pump beam with high spatial and temporal resolution.
This wavepacket is probed by projecting it into a second
(window) wavepacket prepared by the probe \cite{30}. This wavepacket  can be
visualized in real space when the valence excitations are
treated at the configuration-interaction-singles (CIS) level.

\begin{figure}[h]
\centering
\rotatebox{0}{\scalebox{.5}{\includegraphics{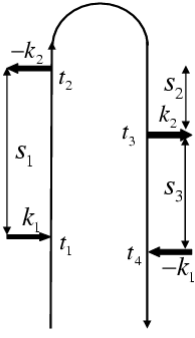}}}\newline
\caption{CTPL diagram representing the RIXS [Eq. (\ref{signal})].
Time runs on the loop clockwise starting from bottom of the left
strand. Interactions on each branch are time-ordered with respect to
each other. Interactions in different branch are not time-ordered.
The incoming and detected modes are denoted by the indices $1$ and
$2$, respectively. $s_1$, $s_2$ and $s_3$ are positive time intervals
along the loop. By ordering the interactions in different branches, Fig. 1
can be decomposed into three  diagrams corresponding to three
different Liouville space pathways \cite{shaul-book}. However, this
will not be necessary here.}
\end{figure}

Equation (\ref{1}) can be derived in the time domain using the closed-time-path loop
(CTPL) diagram shown in Fig. 1. This diagram may then be
modified to represent CXRS. In this diagram time runs clockwise and all
the interactions are ordered along
the loop. In physical time, however, only interactions on the
same branch of the loop are time-ordered with respect to each other; Interactions
on different branches are not time-ordered\cite{christPRA}. The correlation-
function expression for the RIXS signal is readily obtained from
Fig. 1 by assigning each interaction an excitation operator ($B$ or
$B^\dag$) according to the rules given in Ref. \cite{christPRA}. We
then get
\begin{eqnarray}
\label{signal}
&&S_{RIXS}(\omega_1,\omega_2,t) = 2\mbox{Re} \int_{-\infty}^{\infty}ds_2 \int_{0}^{\infty}d s_1
\int_{0}^{\infty}ds_3\nonumber\\
&&  \mbox{e}^{-i\omega_2 s_2} \mbox{e}^{i\omega_1(s_1+s_2+s_3)}E_1(t-s_1-s_2)E^*_1(t-s_3)\nonumber\\
&& \langle B_m(t-s_3)B_m^\dag(t)B_n(t-s_2)B_n^\dag(t-s_1-s_2)\rangle.
\end{eqnarray}
The operator $B^\dag_m$($B_m$) creates (annihilates) a core-hole/valence-electron pair.
\begin{eqnarray}
\label{exciton}
B_m = \sum_{i} \mu_{im}c_i c_m^\dag, ~~ B_m^\dag = \sum_{i} \mu_{mi}c_m c_i^\dag
\end{eqnarray}
Here $c_m$ annihilates electron at the $m$th core orbital and $c_i^\dag$ creates an
electron at the $i$th valence orbital. These are Fermi operators.
$\omega_1$ and $\omega_2$ are the carrier frequencies of the two pulses and
 $\mu_{im}$ is the dipole matrix element between the $i$th valence orbital and $n$th core orbital. $E_j({\bf r},t), j=1,2$ are complex field envelopes
\begin{eqnarray}
E_j({\bf r},t)=E_j\mbox{e}^{ik_j.{\bf r}-i\omega_jt}+ E_j^* \mbox{e}^{-ik_j.{{\bf r}}+i\omega_jt}
\end{eqnarray}
 The time dependence of exciton operators in Eq. (\ref{signal}) is given by the free molecular Hamiltonian (no field), $B_m(t) = \mbox{e}^{iHt}B_m \mbox{e}^{-iHt}$. By inserting the identity operator $\sum_{\nu}|\nu\rangle\langle\nu|$, expanded in the many-body states of the molecule, between the exciton operators in the middle of Eq. (\ref{signal})and assuming stationary field $E_j(t)=1$we can carry out the $s_2$ integration and we immediately recover the Kramers-Heisenberg (KH) expression (\ref{1}) for the signal.

 The single-particle many-body Green's functions provide a standard tool for computing X-ray absorbtion near edge spectra (XANES)\cite{rehr}. Computer codes based on the GW approximation
 developed for these Green's functions are broadly applied to molecules and crystals\cite{ashleyPRB1975,rehrCCR2005,soininenPRB2005,ankudinovPRB1998}.
 We next show that the signal can be approximately expressed in terms of these Green's functions. The correlation function in Eq. (\ref{signal}) can be recast as,
\begin{eqnarray}
\label{r-corr-1}
&&\langle B_m(t-s_3)B_m^\dag(t)B_n(t-s_2)B_n^\dag(t-s_1-s_2)\rangle\nonumber\\
&=&\langle a| B_m U(-s_3)B_m^\dag U(s_2)B_n U(s_1)B_n^\dag|a\rangle
\end{eqnarray}
where $U(s)=\mbox{e}^{iHs}$ is the time evolution operator and $H$
is free molecular Hamiltonian. We set the ground state energy $\omega_a=0$.

We next define the following projection-operator in the
$N+1$-valence electron/1-core-hole space.
\begin{eqnarray}
\label{projection}
{\cal P}_m=\sum_{i}c_i^\dag c_m |a\rangle\langle a|c_m^\dag c_i
\end{eqnarray}
where $|a\rangle$ is the ground many-body state with $N$-valence
electrons. ${\cal P}_m$ selects a sub-space of the full
$N+1$-electron space which includes single valence electron-hole pair
excitations. By inserting the
projection operator (\ref{projection}) twice inside the average on
the r.h.s. of Eq. (\ref{r-corr-1}), we obtain an approximate
expression for the correlation function. Using Eqs.
(\ref{exciton}) and (\ref{projection}), it factorizes into a product of
three correlation functions.
\begin{eqnarray}
\label{r-corr-2}
&&\langle B_m(t-s_3)B_m^\dag(t)B_n(t-s_2)B_n^\dag(t-s_1-s_2)\rangle\nonumber\\
&\approx& \langle a| B_m U(-s_3){\cal P}_m B_m^\dag U(s_2)B_n {\cal P}_n U(s_1)B_n^\dag|a\rangle\nonumber\\
&=& \sum_{ijkl}\sum_{\ip\jp}\mu_{im}\mu_{mk}\mu_{jn}\mu_{nl}\langle a|c_m^\dag c_iU(-s_3)c_\ip^\dag c_m|a\rangle\nonumber\\
&\times& \langle a|c_\ip c_k^\dag U(s_2)c_jc_\jp^\dag|a\rangle \langle a|c_n^\dag c_\jp U(s_1)c_l^\dag c_n|a\rangle.
\end{eqnarray}
\begin{figure}[h]
\centering
\rotatebox{0}{\scalebox{.5}{\includegraphics{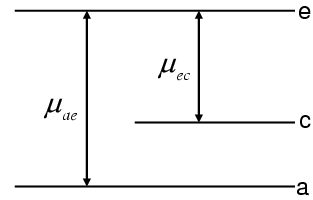}}}\newline
\caption{Three level sequential transition dipole scheme used for pump-probe signal. $a$ and $c$ are valence ground and excited states, and $e$ is the valence $(N+1)$-electron excited state in presence of one core-hole.}
\end{figure}
Since the core-holes are highly localized on the parent
atom, their dynamics is very slow compared to the valence
electrons and may be ignored. We shall ignore this dynamics
and treat the core-hole indices as fixed parameters.

We next introduce the following set of many-body states
\begin{eqnarray}
\label{mb-3} |\chi(i,j)\rangle = c_i c_j^\dag|a\rangle.
\end{eqnarray}
These represent one electron-hole pair excitation state of the
valence $N$-electron system with one electron-hole pair.

We define the one-electron Green's function computed in
the presence of a core-hole at $m$.
\begin{eqnarray}
\label{gf-n}
G_{ij}^{(m)}(t,\tp) = -i\langle T c_i(t)c_j^\dag(\tp)\rangle_m
\end{eqnarray}
where $T$ is the time-ordering operator which rearranges a product
of operators in increasing order in time from the right to the left
and $\langle\cdot\rangle_m$ represents a trace over $N$-electron
ground state of the valence in presence of Coulomb potential due to
a core-hole at $m$.

We also define the one-sided Fourier transform
\begin{eqnarray}
\label{onesided-fourier}
G_{\jp l}^{R(n)}(\omega_1)&=& \int_0^{\infty} dt~ \mbox{e}^{-i\omega_1
t}G_{\jp l}^{(n)}(t)
\end{eqnarray}
where $G^R$ is the retarded Green's functions\cite{fetter}.

Using Eqs. (\ref{mb-3}) and (\ref{gf-n}), the correlation-function (\ref{r-corr-2}) can
be recast as
\begin{eqnarray}
\label{r-corr}
&&\langle B_m(t-s_3)B_m^\dag(t)B_n(t-s_2)B_n^\dag(t-s_1-s_2)\rangle=\nonumber\\
&&\sum_{ijkl}\sum_{\ip\jp}\mu_{im}\mu_{mk}\mu_{jn}\mu_{nl}\theta(s_1)\theta(s_3)\nonumber\\
&&G_{\ip i}^{(m)\dag}(s_3)G_{\jp l}^{(n)}(s_1)\langle \chi(\ip,k)|\mbox{e}^{-iHs_2}|\chi(j,\jp)\rangle.
\end{eqnarray}

The many-body eigenstates $|c\rangle$ (with
energy $\epsilon_c=\omega_{ac}$) of the molecular Hamiltonian
$H$ at the configuration interaction single (CIS) level are given by the linear combinations of single electron-hole pair states.
\begin{eqnarray}
\label{cis}
|c\rangle =\sum_{ij}f_{c;ij}|\chi(i,j)\rangle
\end{eqnarray}
where $f_{c;ij}$ are expansion coefficients. $|c\rangle$
constitute an orthonormal single electron-hole pair basis set,
\begin{eqnarray}
\sum_{c}|c\rangle\langle c|=\sum_{ij}|\chi(i,j)\rangle\langle \chi(i,j)|=1.
\end{eqnarray}
Inserting this identity into the
last term in Eq. (\ref{r-corr}), and substituting it in Eq.
(\ref{signal}) gives
\begin{eqnarray}
\label{final-signal}
&&S_{RIXS}(\omega_1,\omega_2) = \nonumber\\
&&-\sum_{ijkl}\sum_{\ip\jp}\sum_{mnc}
f_{c;\ip k}f_{c;\jp j}^* \mu_{im}\mu_{mk}\mu_{jn}\mu_{nl} \nonumber\\
&&\int_{0}^{\infty}d s_1 \int_{0}^{\infty}ds_3\theta(s_1)\theta(s_3) \mbox{e}^{i\omega_1(s_1+s_3)}\nonumber\\
&&\int_{-\infty}^\infty ds_2 \mbox{e}^{i(\omega_1-\omega_2-\omega_{ac}) s_2}E_1(t-s_1-s_2)E^*_1(t-s_3)\nonumber\\
&&G_{i\ip}^{(m)\dag}(s_3)G_{\jp l}^{(n)}(s_1)
\end{eqnarray}
where $\omega_{ac}=\omega_a-\omega_c$. Equation (\ref{final-signal}) correspods to the RIXS signal obtained earlier using the
Keldysh Green's functions\cite{haug} approach [Eq. 47 in Ref. \cite{privalovPRB64}]. However unlike in Ref. \cite{privalovPRB64},
the RIXS signal derived here takes in to account of the shape of incoming pulse which provides a better control on the signal.

Assuming stationary field envelopes, the $s_2$ integral in Eq. (\ref{final-signal}) reduces to a Dirac-delta function and 
Eq. (\ref{final-signal}) takes the KH form (\ref{1}) with the transition amplitude
\begin{eqnarray}
\label{ampl}
A_{ca}(\omega_1) = \sum_{nlj\jp} f_{c;j\jp}G_{\jp l}^{R(n)}(\omega_1)\mu_{jn}\mu_{nl}
\end{eqnarray}

Equation (\ref{ampl})may be interpreted as follows: The
$N$-electron ground state $|a\rangle$ is excited by the incoming
X-ray beam, creating a core excited state with $(N+1)$ valence
electrons. This state evolves during the short time window permitted by the
core-hole lifetime. This evolution in the presence of a core-hole is
described by the frequency-dependent Green's function. Finally this
excited state is transformed by a dipole transition to a singly
excited valence ground state $|c\rangle$.

\section{X-ray pump-probe simulated raman signal}

We assume high probe intensity so that stimulated emission is dominant
and spontaneous emission can be neglected. The signal is given by
the difference between the transmitted intensities of the probe ($k_2$) with
and without the pump ($k_1$). Four CTPL diagrams contribute to the signal for
our two-band model (Fig. 2).
\begin{figure}[h]
\centering
\rotatebox{0}{\scalebox{.55}{\includegraphics{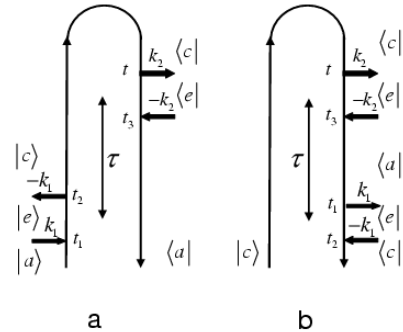}}}\newline
\caption{Two CTPL diagrams that contribute to the pump-probe signal for a two valence level scheme sketched in Fig. 2.}
\end{figure}
When the pump and probe pulse envelopes are well separated in time,
the process may be separated into three steps. First, two interactions
with the pump create a wavepacket of electron-hole pair states $|c\rangle$ in
the valence $N$-electron system. This wave-packet
evolves during the delay $\tau$ and modulates the absorbtion of the
probe. The two loop-diagrams shown in Figs. 3a and 3b contribute to the signal.
These diagrams represent two different correlation functions of the exciton operators.
Details are given in Appendix \ref{appendix-a}.
The signal can be expressed in the Doorway/Window form \cite{IgorPRL,shaul-book,shaul-PRB}.
\begin{eqnarray}
\label{final-signal-111}
S_{CXRS}(\tau) = \sum_{ac}\left[D_{ac}W_{ca}\mbox{e}^{-i\omega_{ac}\tau}+c.c\right].
\end{eqnarray}
The pump prepares a Doorway wavepacket (\ref{dac})
\begin{eqnarray}
\label{100-1}
|D\rangle = \sum_c D_{ac} |c\rangle
\end{eqnarray}
and the probe creates a Window wavepacket (\ref{wca})
\begin{eqnarray}
\label{200-2}
|W\rangle = \sum_c W_{ac} |c\rangle.
\end{eqnarray}
The signal is given by the overlap of these wavepackets. Note that the signal
in (\ref{final-signal-111}) is given for $\tau>0$ (pump interacts before the probe pulse).
However, in order to express the CXRS signal in a form similar to the RIXS (\ref{eq-1}),
for $\tau<0$, we define the CXRS signal to be the same as in Eq. (\ref{final-signal-111}).

Fourier transform of Eq. (\ref{final-signal-111}) gives
\begin{eqnarray}
\label{22}
S_{CXRS}(\omega) &=&  \sum_{ac}D_{ac}W_{ca}\delta(\omega-\omega_{ca})\nonumber\\
&+& \sum_{ac}[D_{ac}W_{ca}]^*\delta(\omega+\omega_{ca}).
\end{eqnarray}

As was done in Eq. (\ref{ampl}) for RIXS, we can express
the Doorway and the Window in terms of the Green's
functions and the CIS coefficients.
Details are given in the Appendix \ref{appendix-b}. We then obtain
\begin{eqnarray}
D_{ac}&:=& P(a) \sum_{ijk}\mu_{im}\mu_{mj}f^*_{c;ik} \int\frac{d\omega}{2\pi}
G_{kj}^{R(m)}(\omega-\omega_{ca})\nonumber\\
&\times& E_1^*(\omega)E_1(\omega-\omega_{ca})\label{eq-1}\\
W_{ca}&:=& 2\sum_{ijk}\mu_{kn}\mu_{nj}f_{c;ji}\int\frac{d\omega}{2\pi}
\mbox{Im}\{G_{ki}^{R(n)}(\omega-\omega_{ca})\}\nonumber\\
&\times& E_2(\omega)E_2^*(\omega-\omega_{ca}).\label{eq-2}
\end{eqnarray}

Note that the exact expression for the pump-probe signal, Eq.
(\ref{final-signal}) with (\ref{dac}) and (\ref{wca}), requires
the computation of many-body states for the $N$ and the
$N+1$ valence electron system.
\begin{figure}[h]
\centering
\rotatebox{0}{\scalebox{.5}{\includegraphics{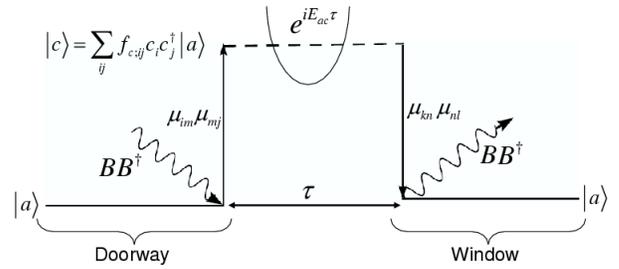}}}\newline
\caption{Schematic representation of Doorway and Window in Eqs. (\ref{eq-1}) and (\ref{eq-2}).}
\label{fig3}
\end{figure}
In contrast, the Green's function expressions, Eq. (\ref{eq-1}) and (\ref{eq-2}),
only require the one-particle Green's functions
and valence excited states at the CIS level. These
may be obtained using standard electronic structure computer codes. In
Appendix (\ref{app-equivelance}) we show that Eqs.
(\ref{dac})-(\ref{wca}) reduce to (\ref{eq-1})-(\ref{eq-2})
within the CIS approximation.

The physical picture offered by the Doorway and the Window in Eq.
(\ref{eq-1}) and (\ref{eq-2}) is shown schematically in Fig. 4.
The system interacts twice with the pump creating a valence excited
state electronic wave packet [Eq. (\ref{100-1}) together with (\ref{eq-1})].
This wave-packet evolves freely for a
time $\tau$, which is not limited by the core-hole lifetime.
The two dipole matrix elements $\mu_{im}$ and $\mu_{mj}$ represent the two
interactions with the X-ray pulse which create and annihilate the
core-hole at $m$. The Green's function $\tilde{G}_{kj}(\omega)$
together with coefficients $f^*{c;ik}$ represents the relevant dynamics.
The Window, Eq. (\ref{200-2}) with (\ref{eq-2}), is similarly created by
the probe X-ray pulse and the signal is given by the overlap of these two wavepackets.

\section{conclusions}

We have presented a many-body Green's function theory of stimulated attosecond
Raman X-ray scattering. The Green's function expression
avoids the explicit computation of the ground and the core-hole
excited states of the system.
The single particle Green's functions can be obtained by
the self-consistent solution of Hedin's equations
\cite{hedin,gunnarsson,onida} which is implemented in standard
computer codes \cite{rehr}. We have assumed that the core-hole is localized and only
enters as a parameter. This  allows to express the signal using single-particle Green's
functions. The "shakeup" and "shakeoff" excitations
\cite{rehr} in which more than one electron is excited due to
creation of a core-hole is included within the CIS
approximation. Equation (\ref{r-corr}) expresses the correlation function of exciton operaotors
in terms of the single-electron Green's functions and the particle-hole propagator for valence orbitals.
The propagator is subsequently expressed in terms of the CIS expansion coefficients for the valence 
excited states. This approximation can be relaxed by calculating this propagator using the time-dependent
Hartree Fock (TDHF). Since TDHF is often insufficient, higher-order corrections can be derived
systematically using the many-body expansion for the particle-hole propagator\cite{lindenberg}.
The present formulation can be extended to
incorporate core-hole and nuclear dynamics through the corresponding
Green's functions using equation of motion technique \cite{lengreth} as was done
in Ref. \cite{privalovPRB64}.

\section*{Acknowledgments}
The support of the Chemical Sciences, Geosciences and Biosciences Division, Office of Basic Energy Sciences, Office of Science and U.S. Department of Energy is gratefully acknowledged. We thank Dr. Igor Schweigert for useful scientific comments.

\appendix

\section{Derivation of Eq. \ref{final-signal-111}}
\label{appendix-a}

The correlation function expression corresponding to the diagrams shown in Figs. 3a and 3b is given by
\begin{eqnarray}
\label{cxrs-1}
&&S_{CXRS}(\tau)= 2\mbox{Re}\int_{-\infty}^\infty dt \int_{-\infty}^t dt_3 \nonumber\\ &\times&\left[\int_{-\infty}^\infty  dt_2 \int_{-\infty}^{t_2} dt_1
\langle\frac{}{} B(t_3)B^\dag(t)B(t_2)B^\dag(t_1)\frac{}{}\rangle \right. \nonumber\\
&+& \left. \int_{-\infty}^\infty  dt_1 \int_{-\infty}^{t_1} dt_2
\langle\frac{}{} B(t_2)B^\dag(t_1)B(t_3)B^\dag(t)\frac{}{}\rangle\frac{}{}\right]\nonumber\\
&&E_1^*(t_2-\tau) E_1(t_1-\tau)E_2(t) E_2^*(t_3).
\end{eqnarray}
Since the pulses are temporally-well-separated and are short compared to the time delay $\tau$, the
upper limits of $t_2$ and $t_3$ integrals in the first and second terms may be
safely extended to infinity. Interchanging the integration variables $t_1$
and $t_2$ in the first term we obtain
\begin{eqnarray}
\label{cxrs}
&&S_{CXRS}(\tau) =\nonumber\\
&& 2\mbox{Re}\int_{-\infty}^\infty dt \int_{-\infty}^t dt_3 \int_{-\infty}^\infty  dt_1 \int_{-\infty}^{t_1} dt_2  E_2(t) E_2^*(t_3)
\nonumber\\
&\times&\left[\frac{}{}\langle\frac{}{} B(t_3)B^\dag(t)B(t_1)B^\dag(t_2)\frac{}{}\rangle E_1^*(t_1-\tau) E_1(t_2-\tau)\right.\nonumber\\
&+&\left.\langle\frac{}{} B(t_2)B^\dag(t_1)B(t_3)B^\dag(t)\frac{}{}\rangle E_1^*(t_2-\tau) E_1(t_1-\tau)\right].\nonumber\\
\end{eqnarray}

The correlation functions in Eq. (\ref{cxrs}) can be
expanded in terms of the many-body states of  the system. By
inserting the identity operator $I=\sum_\nu  |\nu\rangle\langle
\nu|$, where $\nu=c,e$, we obtain
\begin{eqnarray}
\label{corr-n}
&&\langle\frac{}{} B(t_3)B^\dag(t)B(t_1)B^\dag(t_2)\frac{}{}\rangle=\sum_{acee^\prime}P(a)B_{ae}B_{ec}^\dag B_{ce^\prime}B_{e^\prime a}^\dag \nonumber\\
&\times& \mbox{e}^{i\omega_{ec}t} \mbox{e}^{-i\omega_{ea}t_3}
\mbox{e}^{i\omega_{e^\prime a}t_2} \mbox{e}^{-i\omega_{e^\prime c}t_1}
\end{eqnarray}
where $B_{ac}=\langle a|B|c\rangle$, etc. and $P(a)$ is the equilibrium
probability of the system to be in state $|a\rangle$. A similar
expression can be obtained for the other correlation function.
Substituting these results in Eq. (\ref{cxrs}) and defining the
Fourier transform of the pulse envelope as
\begin{eqnarray}
\label{fourier-pulse}
E(t) = \int \frac{d\omega}{2\pi} \mbox{e}^{i\omega t} E(\omega),
\end{eqnarray}
the signal can be expressed in the Doorway/Window representation as \cite{IgorPRL,shaul-book,shaul-PRB}
\begin{eqnarray}
\label{final-signal-1}
S_{CXRS}(\tau) = \sum_{ac}[D_{ac}W_{ca}\mbox{e}^{-i\omega_{ac}\tau}+c.c]
\end{eqnarray}
where the Doorway wavepacket is given by Eq. (\ref{100-1})with
\begin{eqnarray}
\label{dac}
D_{ac}&=& \sum_{e} P(a)B_{ae}B^\dag_{ec}\int \frac{d\omega}{2\pi} \frac{E_1(\omega)E_1^*(\omega+\omega_{ac})}{\omega+\omega_{ec}+i\eta}.
\end{eqnarray}
The Window wavepacket is similarly given by Eq. (\ref{200-2})with
\begin{eqnarray}
\label{wca}
W_{ca} &=& \sum_{e} B_{ce}B^\dag_{ea} \int \frac{d\omega}{2\pi}
\left[\frac{1}{\omega+\omega_{ec}+i\eta}-\frac{1}{\omega+\omega_{ec}-i\eta}\right]\nonumber\\
&&E_2(\omega)E_2^*(\omega+\omega_{ac}).
\end{eqnarray}
Since in the limit $\eta\to 0$
\begin{eqnarray}
\frac{1}{\omega+\omega_{ec}\pm i\eta} = \mbox{PP}\frac{1}{\omega+\omega_{ec}} \mp i\pi \delta(\omega+\omega_{ec}),
\end{eqnarray}
where PP$(1/x)$ denotes the principal part of $1/x$,
the frequency integration in Eq. (\ref{wca}) can be performed, resulting in
\begin{eqnarray}
\label{wcaa}
W_{ca} &=& 2\pi\sum_{e} B_{ce}B^\dag_{ea} E_2(\omega_{ce})E_2^*(\omega_{ac}).
\end{eqnarray}

\section{Green's function expression for the pump-probe signal}
\label{appendix-b}

We start with Eq. (\ref{cxrs-1}) and allow for the creation of a
core-hole at $m$ and $n$. The interaction with $E_1$ ($E_1^*$)
[$E_2$ ($E_2^*$)] creates (destroys) a core-hole at $m$ [n]. We then
have
\begin{eqnarray}
\label{app-1}
&&S_{CXRS}(\tau)= 2\mbox{Re}\int_{-\infty}^\infty dt \int_{-\infty}^t dt_3\nonumber\\ &\times&\left\{\int_{-\infty}^\infty  dt_2 \int_{-\infty}^{t_2} dt_1
\langle\frac{}{} B_n(t_3)B_n^\dag(t)B_m(t_2)B_m^\dag(t_1)\frac{}{}\rangle \right.\nonumber\\
&+& \left.\int_{-\infty}^\infty  dt_1 \int_{-\infty}^{t_1} dt_2
\langle\frac{}{} B_m(t_2)B^\dag_m(t_1)B_n(t_3)B_n^\dag(t)\frac{}{}\rangle
\frac{}{}\right\}\nonumber\\
&\times& E_1^*(t_2-\tau) E_1(t_1-\tau)E_2(t) E_2^*(t_3).
\end{eqnarray}
Inserting the projection operators twice in Eq. (\ref{exciton}), the correlation function in the first term can be expressed
as
\begin{eqnarray}
\label{app-2}
&&\langle\frac{}{} B_n(t_3)B_n^\dag(t)B_m(t_2)B_m^\dag(t_1)\frac{}{}\rangle\nonumber\\
&\approx& \sum_{a}P(a) \langle a|\frac{}{} B_n(t_3){\cal P}_n B_n^\dag(t)B_m(t_2){\cal P}_m B_m^\dag(t_1)\frac{}{}|a \rangle\nonumber\\
&=&\sum_{a}P(a)\sum_{ijkl} \mu_{im}\mu_{mj}\mu_{kn}\mu_{nl} \mbox{e}^{i\omega_a(t_3-t_1)}\nonumber\\
&\times& \langle a| c_kc_n^\dag U(t_3-t){\cal P}_n c_nc_l^\dag U(t-t_2)c_ic_m^\dag {\cal P}_m
U(t_2-t_1)c_mc_j^\dag | a\rangle\nonumber\\
\end{eqnarray}
Substituting this back in Eq. (\ref{app-1}), and changing variables $t_2-t_1=s^\prime$ and $t-t_3=s$, the first term can be
expressed as
\begin{eqnarray}
\label{app-3}
I_1&=& 2\mbox{Re}  \sum_a P(a) \sum_{ijkl}\mu_{im}\mu_{mj}\mu_{kn}\mu_{nl}
\int_{-\infty}^\infty dt \int_0^\infty ds\nonumber\\
&\times& \int_{-\infty}^\infty  dt_2 \int_0^\infty d\sp \mbox{e}^{i\omega_a(t-t_2)}
G_{k\ip}^{(n)\dag}(s) G_{\jp j}^{(m)}(\sp)\nonumber\\
&\times& \langle \chi(\ip,l)|U(t-t_2)| \chi(i,\jp)\rangle\nonumber\\
&\times& E_2(t)E_2^*(t-s) E_1^*(t_2-\tau)E_1(t_2-\sp-\tau)
\end{eqnarray}
where the $|\chi(i,j)\rangle$ basis is defined in Eq. (\ref{mb-3}) and
\begin{eqnarray}
\label{app-4}
\theta(s)\langle a| c_m^\dag c_i U(s)c_j^\dag c_m |a\rangle &=&i\theta(s)\mbox{e}^{-i\omega_as}G_{ij}^{(m)}(s)\nonumber\\
\theta(s)\langle a| c_m^\dag c_i U(-s)c_j^\dag c_m |a\rangle &=&-i\theta(s)\mbox{e}^{i\omega_as}G_{ij}^{(m)\dag}(s).
\end{eqnarray}
Inserting the identity operator in terms of the CIS states in (Eq.
(\ref{cis})), Eq. (\ref{app-3}) finally becomes
\begin{eqnarray}
\label{app-5}
I_1&=& 2\mbox{Re}  \sum_{ac} P(a) \sum_{ijkl}\sum_{\ip\jp}\mu_{im}\mu_{mj}\mu_{kn}\mu_{nl}
f_{c;l\ip}f^*_{c;i\jp}\nonumber\\
&\times&\int_{-\infty}^\infty dt \int_0^\infty ds \mbox{e}^{-i\omega_{ca}t}G_{k\ip}^{(n)\dag}(s)E_2(t)E_2^*(t-s)\nonumber\\
&\times& \int_{-\infty}^\infty dt_2 \int_0^\infty d\sp \mbox{e}^{i\omega_{ca}t_2}
G_{\jp j}^{(m)}(\sp)E_1^*(t_2-\tau)E_1(t_2-\sp-\tau)\nonumber\\
\end{eqnarray}

Performing Fourier transform [Eq. (\ref{fourier-pulse})], Eq.
(\ref{app-3}) can also be expressed as
\begin{eqnarray}
\label{app-6}
I_1&=& 2\mbox{Re}  \sum_{ac} P(a) \sum_{ijkl}\sum_{\ip\jp}\mu_{im}\mu_{mj}\mu_{kn}\mu_{nl}
f_{c;l\ip}f^*_{c;i\jp}\nonumber\\
&\times&\int\frac{d\omega}{2\pi} G_{k\ip}^{R(n)\dag}(\omega-\omega_{ca}) E_2(\omega) E_{2}^*(\omega-\omega_{ca})\mbox{e}^{i\omega_{ca}\tau}\nonumber\\
&\times&\int\frac{d\omega^\prime}{2\pi} G_{\jp j}^{R(m)}(\omega-\omega_{ca}) E_1^*(\omega) E_{1}(\omega-\omega_{ca})
\end{eqnarray}
where $G_{ij}^{R(m)}(\omega)$ is defined in Eq. (\ref{onesided-fourier}).

Proceeding along the same steps that lead from Eq. (\ref{app-2}) to
(\ref{app-6}), and taking the complex conjugate (this is allowed
since we are looking for the real part), the second term in Equation
(\ref{app-1}) can be recast as
\begin{eqnarray}
\label{app-8}
I_2&=& -2\mbox{Re}  \sum_{ac} P(a) \sum_{ijkl}\sum_{\ip\jp}\mu_{im}\mu_{mj}\mu_{kn}\mu_{nl}
f_{c;l\jp}f^*_{c;i\ip}\nonumber\\
&\times&\int\frac{d\omega}{2\pi} G_{k\jp}^{R(n)\dag}(\omega-\omega_{ca}) E_2(\omega) E_{2}^*(\omega-\omega_{ca})\mbox{e}^{i\omega_{ca}\tau}\nonumber\\
&\times&\int\frac{d\omega^\prime}{2\pi} G_{\ip j}^{R(m)\dag}(\omega-\omega_{ca}) E_1^*(\omega) E_{1}(\omega-\omega_{ca}).
\end{eqnarray}

The signal is finally given by
\begin{eqnarray}
S_{CXRS}(\tau) = I_1+I_2
\end{eqnarray}
which results in Eqs. (\ref{eq-1}) and (\ref{eq-2}).

\section{Equivalence of Eqs. (\ref{dac})-(\ref{wca}) and (\ref{eq-1})-(\ref{eq-2}) within the CIS approximation}\label{app-equivelance}

The pump-probe signal is given by Eq. (\ref{final-signal}) where the
Doorway and the Window are given in terms of the many-body states by
Eqs. (\ref{dac}) and (\ref{wca}). Green's function expressions are
given by Eqs. (\ref{eq-1}) and (\ref{eq-2}).

We first expand the GF in the many-body states. Using Eqs.
(\ref{onesided-fourier}) and (\ref{app-4}), we can write
\begin{eqnarray}
\label{app2-4}
&&G^{R(m)}_{ij}(\omega+\omega_{ac})=\nonumber\\&-&i\int_0^\infty dt \mbox{e}^{-i(\omega-\omega_c)t}
\langle a|c_m^\dag c_i U(t) c_j^\dag c_m| a\rangle.
\end{eqnarray}
By inserting the identity operator, $\sum_{e}|e\rangle\langle e|$, before and after the evolution operator and performing the integration over time, Eq. (\ref{app2-4}) reduces to
\begin{eqnarray}
\label{app2-5}
G^{R(m)}_{ij}(\omega+\omega_{ac})=-\sum_e \frac{\langle a|c_i c_m^\dag |e\rangle\langle e|c_m c_j^\dag| a\rangle}{\omega+\omega_{ec}-i\eta}.
\end{eqnarray}
Substituting this in Eqs. (\ref{eq-1}) and (\ref{eq-2}), we obtain
\begin{eqnarray}
D_{ac}&:=& -\sum_{ijke}P(a)\mu_{im}\mu_{mj}f^*_{c;ik}\langle a|c_k c_m^\dag |e\rangle\langle e|c_m c_j^\dag| a\rangle \nonumber\\
&\times&\int\frac{d\omega}{2\pi}
\frac{E_1^*(\omega)E_1(\omega-\omega_{ca})}{\omega+\omega_{ec}-i\eta}\label{app2-6}\\
W_{ca}&:=& -\sum_{ijke}\mu_{kn}\mu_{nj}f_{c;ji} \langle a|c_i c_n^\dag |e\rangle\langle e|c_n c_k^\dag| a\rangle\nonumber\\
&\times&\int\frac{d\omega}{2\pi}
\left[\frac{1}{\omega+\omega_{ec}+i\eta}-\frac{1}{\omega+\omega_{ec}-i\eta}\right]\nonumber\\
&\times&E_2(\omega)E_2^*(\omega-\omega_{ca}).\label{app2-7}
\end{eqnarray}

We next consider the expressions for the Doorway and Window in terms of the many-body states.
Using Eq. (\ref{exciton}), Eqs. (\ref{dac}) and (\ref{wca}) can be expressed as,
\begin{eqnarray}
\label{dac-1}
D_{ac}&=& \sum_{ije} P(a)\mu_{im}\mu_{mj}\langle a|c_i c_m^\dag |e\rangle
\langle e|c_m c_j^\dag |c\rangle \nonumber\\
&&\int \frac{d\omega}{2\pi} \frac{E_1(\omega)E_1^*(\omega+\omega_{ac})}{\omega+\omega_{ec}+i\eta}
\end{eqnarray}
and
\begin{eqnarray}
\label{wca-1}
W_{ca} &=& \sum_{ije} \mu_{kn}\mu_{nj}  \langle c|c_k c_n^\dag |e\rangle
\langle e|c_n c_j^\dag |a\rangle\nonumber\\
&& \int \frac{d\omega}{2\pi}
\left[\frac{1}{\omega+\omega_{ec}+i\eta}-\frac{1}{\omega+\omega_{ec}-i\eta}\right]\nonumber\\
&&E_2(\omega)E_2^*(\omega+\omega_{ac}).
\end{eqnarray}
Using the CIS expansion for the excited states $|c\rangle$, Eq.
(\ref{cis}), we can write
\begin{eqnarray}
\label{nn-1}
\langle e|c_m c_j^\dag |c\rangle &=& \sum_{\ip\jp} f_{c;\ip\jp}
\langle e|c_m c_j^\dag c_{\ip}c_{\jp}^\dag|a\rangle\nonumber\\
&=& \sum_{\jp} f_{c;j\jp} \langle e|c_m c_{\jp}^\dag|a\rangle
\end{eqnarray}
where in going to the second line the sum over $\ip$ can be done
since state $|e\rangle$ corresponds to single excited $N+1$-valence
electrons. Substituting (\ref{nn-1}) in (\ref{dac-1}) we get
\begin{eqnarray}
\label{dac-2}
D_{ac}&=& \sum_{ijke} P(a)\mu_{im}\mu_{mj}f_{c;jk}\langle a|c_i c_m^\dag |e\rangle
\langle e|c_m c_k^\dag |a\rangle \nonumber\\
&&\int \frac{d\omega}{2\pi} \frac{E_1(\omega)E_1^*(\omega+\omega_{ac})}{\omega+\omega_{ec}+i\eta}
\end{eqnarray}
Taking the complex conjugate and interchanging dummy indices $i$ and
$j$, Eq. (\ref{dac-2}) becomes same as (\ref{app2-6}). Similarly one
can show that Eq. (\ref{wca-1}) is same as (\ref{app2-7}) within CIS
approximation.

This proves the equivalence of Eqs. (\ref{dac})-(\ref{wca}) and Eqs. (\ref{eq-1})-(\ref{eq-2}).

\end{document}